# Impact of Public and Private Investments on Economic Growth of Developing Countries

Faruque Ahamed[1]

**Abstract**

This paper aims to study the impact of public and private investments on the economic growth of developing countries. The study uses the panel data of 39 developing countries covering the periods 1990-2019. The study was based on the neoclassical growth models or exogenous growth models state in which land, labor, capital accumulation, etc., and technology proved substantial for economic growth. The paper finds that public investment has a strong positive impact on economic growth than private investment. Gross capital formation, labor growth, and government final consumption expenditure were found significant in explaining the economic growth. Overall, both public and private investments are substantial for the economic growth and development of developing countries.

## Introduction

In the last few decades, economists are trying to understand the global and country-specific factors contributing to economic growth. The variations in market fluctuations, financial crises, economic turmoil, and recessions made it hard to predict the economic uncertainty. An array of dynamic macroeconomic factors has been analyzed to determine a robust framework and effective model to explain the preceding performances and forecast future behavior of the economies. Both developed and developing countries face the change in macroeconomic dynamics and identified a complex relationship between public & private investment with economic growth. Public and private investments play a substantial role in the production functions by providing the required capital for development. Public investment may create scope for private investment through resource creation and socio-economic infrastructure support. However, Public investment can have a crowding-out effect on private investment. The IS-LM theory clearly states the impact of public investment on private investment. An increase in government spending, taxes, and domestic interest rate can lead to a parallel shift in the IS curve and adversely affect private investment. (Buiter, 1977; Ram, 1986). Public investment may also deprive private investment by competing in investment goods and utilizing physical and financial resources. Empirical studies by Aschauer (1989), Seitz (1994), and Pereira (2001) revealed that public investment is crowding in private investment, whereas Zou (2003) finds that public investment has a crowding-out effect.

Researchers are yet to find a conclusive idea about the public and private investment collaboration and economic growth in the above circumstances. Public investments are working as business stimulus policy for almost the last three decades. Public capital enhances the productivity of the private capital and drives the return on investment upward.

[1] Department of Economics, Northern Illinois University.

# Theoretical and Empirical Review

*Economic Growth Theory*

Researchers have identified two Economic growth models: The neo-classical growth model and the New growth model. Neo-classical growth models or exogenous growth models state that factors of production such as land, labor, capital accumulation, etc., and technological variables affect long-run productivity and economic growth (Solow, 1956). The new growth model or endogenous growth model argues that innovation and public-private investments in human capital fuel economic growth. (Romer, 1996; Lucas, 1988; Barro, 1990; Rebelo, 1991). Apart from these models, researchers also used Cobb-Douglas production function (1928) to identify the effect of labor and capital on output. (Aschauer, 1989). Most of the models confirm that public investment influences economic growth affecting aggregate demand and aggregate supply.

Let's consider, there is an representative agent in the economy wants to maximize the utility with the following function:

$$V = \sum_{t=1}^{\infty} \delta^{t-1} u(c_t, l_t g_t)$$

In the equation $\delta$ is the discount factor where $0 < \delta < 1$, $c$ is the consumption, $w$ is the work effort and $g$ implies the total government expenditure. Consumption and work effort is associated with the utility and disutility. It is assumed that government expenditure connects with private utility through private consumption goods or leisure activities. The representative agent can draw the production function with constant return to scale as follows:

$$Y = F_t(L, K, G) = A L^\alpha K^\beta G^\gamma \qquad (1)$$

Where, L is the labor, K is private contribution in capital stock, G is government contribution in capital stock and A is technical coefficient.

Case 1: Lets consider $\alpha+\beta=1$ where total output is allocated to its labor and private capital factors. In this case, government contribution in capital stock affects the other two factors. When the public capital works as supplement for the private capital stock, affecting the marginal productivity of private capital stock through inducing private investment and output. The marginal productivity of labor also faces positive impact as both public and private capital per worker increases. Taking the derivates of labor and private capital in Equation (1) we get,

$$Y = \frac{dF}{dL}L + \frac{dF}{dk}K$$
$$= (\alpha A L^{\alpha-1} K^\beta G^\gamma)L + (\beta A L^\alpha K^{\beta-1} G^\gamma)K$$
$$= (\alpha \frac{Y}{L})L + (\beta \frac{Y}{K})K$$
$$= MPL*L + MPK*K$$

Until the marginal productivity of labor and marginal productivity of private capital is equal there will be adjustments of labor and capital.

Case 1: Now, Let's consider $\alpha+\beta+\gamma=1$ where government investments/capital is an independent factor like the other two factors. When the public capital stock is independent and complements the private capital stock and labor then an increase in public capital directly affects the output considering the *Ceteris Paribus* holds. Taking the derivates of labor, Private capital and public capital stock in Equation (1) we get,

$$Y = \frac{dF}{dL}L + \frac{dF}{dk}K + \frac{dF}{dG}G$$
$$= (\alpha AL^{\alpha-1}K^{\beta}G^{\gamma})L + (\beta AL^{\alpha}K^{\beta-1}G^{\gamma})K + (\gamma AL^{\alpha}K^{\beta}G^{\gamma-1})G$$
$$= (\alpha \frac{Y}{L})L + (\beta \frac{Y}{K})K + (\gamma \frac{Y}{G})G$$
$$= MPL*L + MPK*K + MPG*G$$

Until the marginal productivities of public capital, labor and private capital are equal there will be adjustments of labor and capital(s) to reach an equilibrium.
Placing the value $\alpha = 1-\beta-\gamma$, we get

$$\ln \frac{Y}{L} = \ln A + \beta \ln \frac{K}{L} + \gamma \ln \frac{G}{L}$$

Setting $a = \ln A$, $y = Y/L$, $k = K/L$ and $g = G/L$ we can write

$$\ln y = a + \beta \ln k + \gamma \ln g$$

By adjusting two periods, *t* and *t-1*, we can obtain,

$$\ln (y_t/y_{t-1}) = a + \beta \ln (k_t/k_{t-1}) + \gamma \ln (g_t/g_{t-1})$$

Considering constant population growth rate,

$$\ln (Y_t/Y_{t-1}) = a + \beta \ln (K_t/K_{t-1}) + \gamma \ln (G_t/G_{t-1})$$

The last two equations suggests that there exists direct link between publication and private investment with economic growth.

*Empirical Studies*
Researchers are working to investigate the relation of public and private investment with economic growth for many years. The studies, however, revealed different results depending on the sample and method used.
Le and Suruga (2005) investigate the impact of public investment and foreign direct investment (FDI) on economic growth using the panel data of 105 developed and developing countries for 1970-2009. The results revealed that both public investment and FDI have a positive impact on GDP. The results also showed that when the public investment exceeds the threshold of 9%, the effect of FDI on economic growth becomes weaker, and the research concludes that excessive public investment erases the benefit from FDI. Blejer and Khan (1984) explore the crowding out or crowding in public investment on private investment for 24 developing countries from 1971-1979. They identified that public investment in infrastructure is beneficial for private investment, whereas another type dry fund of private investment. Ashauer (1989) examines whether public investment crow out private investment using the United States data for 1925-1985. The results show that public investment increases the marginal productivity of the private capital, and higher public spending reduces private investment. The study found both the crowd-in and crowd-out effect of public investment. Everhart and Sumlinski (2001) tried to find the partial correlation between public and private investment using a data panel of 63 developing countries over the period 1970-2000. The studies conclude that there exists a negative correlation between public and private investment. However, the correlation seems to be positive for the countries with a better institutional framework.

Deverajan et al. (1996) explore the relationship between public expenditure and economic growth using a sample of 43 developed and developing countries over 1970-1990. The study finds that public capital expenditure has a positive effect on the economic growth of developing countries, whereas the effect is negative for developing countries. The excessive allocation of resources by public investment can become unproductive and inefficient for the economy. Ghosh and Gregoriou (2007) also conclude similar results in developing countries' optimal fiscal policy framework.

Barro (1990) examines the relationship between the growth rate of per real capita GDP and share of government expenditure using the extended endogenous growth model for 76 countries from 1960 to 1985. He concluded that government consumption is inversely related to economic growth, but the relationship between public investment and economic growth is found to be insignificant. Hsieh and Lai (1994) used a similar model to analyze the relationship among the growth rate of per capita GDP, the ratio of private investment to GDP, and government expenditure for G-7 countries. The study found that government spending has a significant impact on the growth rate of per capita GDP for Canada, the UK, and Japan but an insignificant impact for France, Germany, Italy, and the USA. However, the private investment to GDP ratio has a significant effect on the USA, Japan, Canada, Germany, and the UK. Zou (2006) performs a study on the interaction between public and private investment and economic growth for the USA and Japan. He suggests that both public and private investment have a significant contribution to Japanese economic growth. For the USA, the private investment seems to play a much more significant role than public investment.

Ghali (1998) conducts a study in Tunisia, in which IMF implemented a debt-stabilization program and identified that public investment has a long-run inverse effect on economic growth. Ramirez and Nazmi (2003) made an empirical study for nine Latin American countries and concluded that both public and private investment positively impact GDP growth. Phetsavong and Ichihashi (2012) investigate the impact of FDI, public investment, and private domestic investment using a sample of 15 Asian developing countries from 1984 to 2009. They suggest that private domestic investment and FDI are the two most crucial contributing factors for economic growth, whereas public consumption is inversely related to economic growth. The studies also found that public investment reduces the positive impact of private domestic investment and FDI on economic growth. Nguyen and Trinh (2018) examined both short and long-term influences of public investment on private investment and economic growth. The authors use an autoregressive distributed lag model using Vietnam's macro data from 1990-2016. The results indicate that public investment has a positive effect on short-term and negative effects in constraining long-term growth. However, private investment and FDI have positive effects on short-term economic growth. State-owned capital stock has positive impacts on economic growth in both the short and long run. Ahamed (2021) suggests that unexpected events like Covid-19 can also impact the private and public investments, thereby hurting the economic growth.

## Data and Methodology

*Data*

The study uses panel data of 39 developing countries throughout 1990-2019 which comprises 7020 observations. The data has been collected from mainly two sources: World Development Indicator and the International Monetary Fund database. Sample data from developing countries across all the continents are chosen to balance the research data. Domestic private credit and foreign direct investment are used as a proxy of private investments, whereas the gross capital formation, labor growth, and government consumption expenditures are considered public investments.

*Table 1: Summary statistics of the variables*

| Variable Description | Mean (Std. Dev.) | Min | Max |
| --- | --- | --- | --- |
| *Dependent Variable* | | | |
| GDP Growth Rate | 3.9367 (4.4138) | -50.25 | 35.22 |
| *Variables of interest* | | | |
| Domestic Private Credit (% of GDP) | 30.6404 (31.6157) | 1.6155 | 160.1248 |
| Foreign Direct Investment (% of GDP) | 2.4002 (3.7685) | 8.7031 | 46.2752 |
| Gross Capital Formation (% of GDP) | 22.1185 (8.6896) | 1.5252 | 77.8900 |
| Labor Growth Rate | 0.0261 (0.0168) | -0.0449 | 0.1208 |
| Government Consumption Expenditure (% of GDP) | 13.7099 (4.9718) | 0.9112 | 31.5544 |
| No. of observations | 7020 | | |

*Empirical Model & Tests*

As the dataset for this study is a panel data set, to examine the research question of whether the public and private investments affect the economic growth, a fixed-effect or a random-effect model can be used to control for time-invariant un-observables that affect both dependent and key independent variables. In particular, the fixed and random-effect models incorporate an individual-specific time-invariant factor, $\alpha_i$. If it is certain that, is not correlated with all independent variables and is normally distributed, then the random effects model would be appropriate. However, in this case, it not certain that $\alpha_i$ is not correlated with an independent variable. The fixed-effect model is used in this study to avoid the stronger assumption required in this case of the random-effect model. The choice of model is also consistent with the Hausman (1978) specification test.

A two-way (country and year) fixed-effect model comprised of both endogenous and exogenous variables is used as a first step. In the second step, Pooled Ordinary Least Squared (POLS) has been used to check the robustness of the model. The model is described below:

$$GDP_{ij} = \alpha_i + \gamma_t + \delta_1 DC_{ij} + \delta_2 FDI_{ij} + \delta_3 CAP_{ij} + \delta_4 LAB_{ij} + \delta_5 CON_{ij} + \epsilon_{ij}$$

where,

$GDP_{ij}$ = Economic Growth Rate
$\alpha_i$ = Country specific time-invariant factor
$\gamma_t$ = Year-specific fixed effects
$DC_{ij}$ = Total domestic credit (% of GDP)
$FDI_{ij}$ = Foreign Direct Investment, Net Inflows (% of GDP)
$CAP_{ij}$ = Gross Capital Formation (% of GDP)
$LAB_{ij}$ = Labor Growth Rate
$CON_{ij}$ = Government Final Consumption Expenditure (% of GDP)

In the estimation, *i* and *t* denote for the number of cross section country (1,2,3…N) and time (1990,1991….2019) respectively.

## Empirical Results and Analysis

*Results*

As the data set is a panel, the Hausman test was conducted to identify the fixed-effects model or a random-effects model. The tests examine the correlation between the unique errors and the regressors in the model. The null hypothesis states that there is no correlation and preferers a random-effects model. The estimated p-value is 00001388, so the null hypothesis has been rejected. It is concluded that the fixed-effect model is the preferred model for this study.

Both fixed-effect and ordinary least square methods exhibit consistency in the results. Domestic private credit has a positive but insignificant effect on economic growth. Foreign direct investment is inversely related to economic growth, yet the effect is insignificant. The private investment variables are showing mixed results in affecting the GDP growth. The gross capital formation is found to be positive and significantly related to economic growth. Another factor that has a statistically significant and positive relationship with economic growth is the labor growth rate. In both the fixed effect and OLS model, one unit increase in labor growth increases the economic growth of the developing countries 21.2778 and 20.1560 percentage points. The government consumption expenditure has a negative yet significant relationship with GDP.

*Table 2: Empirical results*

| Dependent Variable | (1) | (2) |
|---|---|---|
| GDP Growth Rate | Fixed-effect model | Ordinary Least Square |
| *Variables of interest* | | |
| Domestic Private Credit (% of GDP) | 0.00453 | 0.004711 |
| | (0.00410) | (0.004153) |
| Foreign Direct Investment, New Inflows (% of GDP) | -0.04079 | -0.014589 |
| | (0.03445) | (0.034461) |
| Gross Capital Formation (% of GDP) | 0.12488*** | 0.119190*** |
| | (0.01523) | (0.015313) |
| Labor Growth Rate | 21.2778** | 20.1560** |
| | (7.35090) | (7.451923) |
| Government Consumption Expenditure (% of GDP) | -0.10605*** | -0.120566*** |
| | (0.02529) | (0.025651) |
| Number of observations | 7020 | 7020 |
| R-sq | 0.084 | 0.082 |
| Adjusted R-sq | 0.057 | 0.078 |

**5% Significance level
***1% Significance level

*Table 2: Correlation Matrix*

| | GDP | DC | FDI | CAP | LAB | CON |
|---|---|---|---|---|---|---|
| GDP | 1.000000 | | | | | |
| DC | 0.049859 | 1.000000 | | | | |
| FDI | 0.058018 | 0.098937 | 1.000000 | | | |
| CAP | 0.245537 | 0.222324 | 0.291234 | 1.000000 | | |
| LAB | 0.066328 | -0.1158 | 0.040662 | 0.015698 | 1.000000 | |
| CON | -0.13415 | 0.190755 | 0.031906 | -0.04275 | 0.073264 | 1.000000 |

*Analysis*

A general conception is public and private investment induces economic growth and creates development opportunities. However, previous researchers suggest the situation varies depending on various circumstances of investments. Developing countries lack infrastructural capital stock and face weak budgetary positions but possess a strong labor force, so the magnitude of public investment affecting economic growth is higher than the private investment. In developed countries, however, private investment seems to have a higher effect due to strong infrastructure and higher private capital formation. Although government investment in infrastructure, etc., sectors are crucial, excessive public investment in developing countries can have a crowding-out effect on the private investment. The disparity of public-private investment rates of return creates disparity and discourages private domestic investors.

Public investment may influence private investment and boost the economy for the short term, but it may have significant adverse effects in the long term. As the government budgetary decisions in developing countries are inefficient and unfocused, overspending may require borrowing from private sources. State-owned enterprises lack transparency and accountability, which hinders the plan of creating clean and sustainable development. Developing countries have a greater population growth rate and an increased workforce. Still, due to poor quality education and inadequate investment in education, the majority of the workforce stays in unskilled territory. However, excessive public investment may cause the vitality of the economy to be lost and make private investment less productive in the long run. Transition to developed countries from developing countries requires policymakers to prioritize private investments and design public policies to stimulate private investment. Public investment works as a complementary service for domestic private investment and foreign direct investments. Developing countries follow public investment-driven economies in terms of sustainable economic growth and exert a strong influence over the private sector. Realizing the importance of private investment, China has empowered private sectors through synchronization with the public policies and investment to create a robust bidirectional relationship. This private-public investment correlation triggers economic growth and fosters economic development.

The Neoclassical growth theory implies that a steady economic growth rate results from three driving forces: labor, capital, and technology. Capital accumulation and technological investment are the critical challenges faced by developing countries. Limited resource availability can only be offset by technological innovation considering capital and labor productivity. The main instrument of economic growth varies at various stages of a country's development. Physical capital accumulation and human capital development are substantial for early development stages but maintaining steady growth investments in technological research and development is vital. These are lacking in developing countries that fail private investments and discourages foreign investments. Lack of policy, skilled human resources, and infrastructure support the determinants of economic growth variables that fail to generate interactive relationships.

## Conclusion

The study uses panel data from 39 developing countries during the period 1990-2019. The effect of public and private investments in determining economic growth rate has been explored in the study. The results of this paper suggest some implications in constructing a theoretical model which structures the impact of public-private investments on economic growth. The empirical results show that private investment has mixed effects (mostly positive) on GDP growth. Domestic private credit has a positive, whereas foreign direct investment has a slightly negative impact on economic growth. The labor force growth rate drives economic growth significantly. Gross capital formation is associated with higher economic growth, while the government consumption expenditure is inversely related to the growth. Most of the countries in the study focus on public investment-driven growth and try to exert control over private investments. Focus on human capital development is substantial for economic growth, but the countries have less concentration on that sector. The evidence suggests the countries need to improve the public sector productivity and develop an analytical framework to stimulate private investment. Policy support to facilitate the private investment and skilled labor force can ensure a stable macroeconomic environment and sustainable economic growth.